\documentclass[12pt]{article}

\newcommand\DD {{\cal D}}

\newtheorem{theorem}{Theorem}[section]

\newcommand{\eq}[1]{Eq.~(\ref{#1})}
\newcommand{\beq}{\begin{equation}}
\newcommand{\eeq}{\end{equation}}
\newcommand{\bea}{\begin{eqnarray}}
\newcommand{\eea}{\end{eqnarray}}
\renewcommand{\and}{{\qquad {\rm and} \qquad}}
\newcommand{\where}{{\qquad {\rm where} \qquad}}

\newcommand{\virg}{{\qquad , \qquad}}
\newcommand{\tr}{{\,\rm tr}\:}
\newcommand{\ee}[1]{{{\rm e}^{#1}}}
\renewcommand{\d}{{{\partial}}}
\newcommand{\D}{{{\hbox{d}}}}

\begin{document}
\sloppy


\pagestyle{empty}
\hfill SPHT01/092 CRM 2755
\addtolength{\baselineskip}{0.20\baselineskip}
\begin{center}
\vspace{26pt}
{\large \bf {A concise expression for the ODE's of orthogonal polynomials.}}
\newline
\vspace{26pt}

{\sl B.\ Eynard}\hspace*{0.05cm} $^{\dagger}$ $^{\ddagger}$\footnote{ E-mail: eynard@spht.saclay.cea.fr }\\
\vspace{6pt}
$^{\dagger}$Service de Physique Th\'{e}orique de Saclay,\\
F-91191 Gif-sur-Yvette Cedex, France.\\
$^{\ddagger}$Centre de recherches math\'ematiques,
Universit\'e de Montr\'eal\\
C.~P.~6128, succ. centre ville, Montr\'eal,
Qu\'ebec, Canada H3C 3J7.\
\end{center}

\vspace{20pt}
\begin{center}
{\bf Abstract:}
\end{center}
%


It is known that orthogonal polynomials obey a $3$ terms recursion relation, as well as a $2\times 2$ differential system.
Here, we give an explicit representation of the differential system in terms of the recursion coefficients.
This result is a generalization of an expression of Fokas, Its and Kitaev for the symmetric case.




\newpage
\pagestyle{plain}
\setcounter{page}{1}


\section{Introduction}

Orthogonal polynomials\cite{Szego} continue to be extensively studied at the moment, partly because of their intrinsic interesting structures, and also because of their relationship with many physical and mathematical applications.
For example they are the main tool for studying the large $N$ asymptotics of the spectral statistics of random matrices \cite{Mehtaorthpol, Mehta, BI, ZJDFG, Guhr, courseynard, TW2}.

Orthogonal polynomials are known to satisfy 3-term recurrence relations, as well as differential equations.
These differential equations are very useful for the derivation of the large $N$ asymptotics, and the associated Riemann-Hilbert problem \cite{BI}.

For any given weight of the form $\ee{-V(x)}$, where $V(x)$ is a polynomial, finding the differential system is easy, but so far, no general expression was known (except when $V(x)$ is an even polynomial \cite{FIK}). The calculations are straightforward but more and more involved as the degree of $V$ increases.
Here, we derive an explicit and concise expression, valid for any polynomial $V(x)$.
The expression is very similar to that of \cite{FIK}.

In addition, these orthogonal polynomials satisfy differential equations with respect to the coefficients of $V(x)$, viewed as deformation parameters.
Here, we derive an explicit expression for these deformation equations. 
 
\bigskip

{\noindent \bf Outline of the article:}

- In part 2, we introduce the orthogonal polynomials and set the notations.

- In part 3, we give the expression of the differential system satisfied by the orthogonal polynomials.

- In part 4, we give the expression of the deformation equations satisfied by the orthogonal polynomials, and give an alternative proof of the results of part 2.

\section{Orthogonal polynomials}

Consider the family of monic polynomials $p_n(x)=x^n+\dots$, orthogonal with
respect to the weight $\ee{-V(x)}$:
\beq
\int_{\Gamma} \D{x}\, p_n(x) p_m(x) \,\,\ee{-V(x)} = h_n \delta_{nm} \ ,
\eeq
where $V(x)$ is referred to as the potential.
The orthogonal polynomials exist under suitable asumptions on $V$, and are unique \cite{mcl, Szego}.

For simplicity, we will assume that $V$ is a polynomial of even degree, with positive leading coefficient, and the the integration contour $\Gamma$ is the real axis\footnote{Many of these asumptions can easily be lifted, see section 4.}.

Instead of the $p_n$'s, it is more convenient to introduce the following
quasi-polynomial functions:
\beq\label{defpsin}
\psi_n(x) = {1\over \sqrt{h_n}}\, p_n(x)\, \ee{-{1\over 2} V(x)} \ ,
\eeq
which are orthonormal:
\beq\label{ortho}
\int \D{x}\, \psi_n(x) \psi_m(x) = \delta_{nm} \ .
\eeq

\subsubsection{Recursion equation}

It is well known \cite{Szego, Mehta} that the orthogonal polynomials satisfy a three-term recursion relation, which reads:
\beq\label{multx}
x \psi_n(x) = \sum_{m=0}^\infty Q_{n,m} \psi_m  = \gamma_{n+1} \psi_{n+1} + \beta_n \psi_n + \gamma_n \psi_{n-1} \ ,
\eeq
where 
\beq\label{gammahn}
\gamma_n = Q_{n,n-1} = \sqrt{h_{n}\over h_{n-1}} \and \beta_n = Q_{n,n}  \ .
\eeq
The infinite matrix $Q$ is symmetric ($Q_{n,m}=Q_{m,n}$), and has only 3
bands ($Q_{m,n}=0$ if $|m-n|>1$):
\beq\label{defQ}
Q = \pmatrix{\beta_0 & \gamma_1 & 0 & \dots & \dots \cr
             \gamma_1 & \beta_1 & \gamma_2 & 0 & \cr
             0 & \gamma_2 & \beta_2 & \gamma_3 & \cr
             \vdots & 0 & \ddots & \ddots & \ddots 
} \ .
\eeq

\subsubsection{The string equation}

The derivative of $\psi_n$ can be expanded on the basis $\left\{ \psi_m \right\}_{m=0..\infty}$:
\beq\label{derivx}
\psi_n'(x) = \sum_{m=0}^\infty P_{n,m} \psi_m(x) \ ,
\eeq
Integrating \eq{ortho} by parts implies that the infinite matrix $P$ is
antisymmetric: $P_{n,m}= - {P_{m,n}} $, and from $p_n' = n p_{n-1} +
O(x^{n-2})$, one has:
\bea
P_{n,m} + {1\over 2} \left(V'(Q)\right)_{n,m} & = & 0 \qquad {\rm if} \,\, m\geq n  \\
P_{n,n-1} + {1\over 2} \left(V'(Q)\right)_{n,n-1} & = & {n\over \gamma_n} \ ,
\eea
which implies
\bea\label{eqP}
P_{n,m} & = & -  {1\over 2} \left(V'(Q)\right)_{n,m}  \qquad {\rm if} \,\, m\geq n  \\
P_{n,m} & = & +  {1\over 2} \left(V'(Q)\right)_{n,m}  \qquad {\rm if} \,\, m\leq n \ .
\eea
This can be written:
\beq\label{PV}
P = -{1\over 2} \left( V'(Q)_+ - V'(Q)_- \right) \ ,
\eeq
where for any infinite matrix $A$, $A_+$ (resp. $A_-$), denotes the upper (resp. lower) triangular part of $A$.

In particular, \eq{eqP} implies the following equations, known as ``string equations'':
\beq\label{eqmvt}
0  =  V'(Q)_{n,n}  \virg {n\over \gamma_n}   =  V'(Q)_{n,n-1}\label{eqVg} 
\eeq

\subsection{Differential system}

The sum in \eq{derivx} contains only a finite number of terms and can be written:
\beq\label{psinprimeeta}
\psi_n'(x) = -{1\over 2} \sum_{k=1}^{{\rm deg} V'} \eta_{k,n} \psi_{n+k} - \eta_{k,n-k} \psi_{n-k}
\where
\eta_{k,n} = V'(Q)_{n,n+k} \ .
\eeq
Note that
\beq
V'(x)\psi_n(x) =  \sum_{k=1}^{{\rm deg} V'} \eta_{k,n} \psi_{n+k} + \eta_{k,n-k} \psi_{n-k} \ .
\eeq

Using \eq{multx} recursively, one can write any $\psi_{n+k}$ with $k>0$ and any $\psi_{n-1-k}$ with $k>0$ as a linear combination of $\psi_n$ and $\psi_{n-1}$ with coefficients polynomial in $x$, of degree $k$.
\eq{psinprimeeta} can thus be rewritten as a $2\times 2$ differential system:
\beq
{\D\over \D{x}} \pmatrix{\psi_{n-1}(x) \cr \psi_{n}(x)} = \DD_n (x)
\pmatrix{\psi_{n-1}(x) \cr \psi_{n}(x)} 
\eeq
where $\DD_n(x)$ is a $2\times 2$ matrix with polynomial coefficients of degree
at most $d=\deg V'$.
It is known that it must have the form \cite{Szego}:
\beq
\DD_n (x) = {1\over 2} V'(x) \pmatrix{1 & 0 \cr 0 & -1} + \pmatrix{ \gamma_n u_n(x) & -\gamma_n v_{n-1}(x)
\cr \gamma_n v_{n}(x) &   - \gamma_n u_n(x)  } 
\eeq
where $u_n(x)$ and $v_n(x)$ are polynomials of degrees $\deg u_n(x) \leq d-2$ and $\deg v_n(x) \leq d-1$.
What was not known so far, is how to express $u_n(x)$ and $v_n(x)$ in terms of the coefficients $\gamma_n$ and $\beta_n$ (i.e. the coefficients of $Q$), apart form the case where $V$ is an even polynomial \cite{FIK}.
Here, we prove the following theorem:

\begin{theorem}\label{thDeriv}

The matrix $\DD_n(x)$ is:
\beq\label{thDD}
\DD_n(x) =  {1\over 2} V'(x)  \pmatrix{1 & 0 \cr 0 & -1} + \pmatrix{ \left( {V'(Q)-V'(x)\over Q-x}\right)_{n-1,n-1} & \left( {V'(Q)-V'(x)\over Q-x}\right)_{n-1,n}
\cr \left( {V'(Q)-V'(x)\over Q-x}\right)_{n,n-1} &   \left( {V'(Q)-V'(x)\over Q-x}\right)_{n,n}  } \, \pmatrix{ 0 & -\gamma_n \cr \gamma_n & 0 }  
\eeq
\end{theorem}

\noindent{\bf Proof:} \\
One can check from $[\D/ \D{x},x]=1$ and from the compatibility with the shift ($n\to n+1$), that $u_n(x)$ and $v_n(x)$ must satisfy the following recurrences:
\bea
0 & = & V'(x) + \gamma_n u_n(x) + \gamma_{n+1} u_{n+1}(x) + (\beta_n-x)v_n(x) \label{recuva}\\
1 & = & (\beta_n-x)(\gamma_{n+1}u_{n+1}(x)-\gamma_n u_n(x)) + \gamma_{n+1}^2 v_{n+1}(x) - \gamma_n^2 v_{n-1}(x) \label{recuvb} \ .
\eea
and these recurrences uniquely determine $u_n(x)$ and $v_n(x)$ for all $n$, provided that the initial terms are given.

Now, consider the infinite matrix:
\beq
R(x) = \left({V'(Q)-V'(x)\over Q-x}\right) \ .
\eeq
Notice it is symmetric $R_{n,m}=R_{m,n}$.

Computing the diagonal $(n,n)$ term of $(Q-x)R(x)$ and using \eq{eqmvt} gives:
\beq
V'(Q)_{n,n} - V'(x) = - V'(x) = (\beta_n-x) R_{n,n}(x)  + \gamma_{n} R_{n-1,n}(x) +\gamma_{n+1} R_{n+1,n}(x) \,
\eeq
and computing the $(n,n\pm 1)$ terms, using \eq{eqmvt} gives:
\beq
V'(Q)_{n,n+1}  = {n+1\over \gamma_{n+1}} = (\beta_{n}-x) R_{n,n+1}(x)  + \gamma_{n} R_{n-1,n+1}(x) + \gamma_{n+1} R_{n+1,n+1}(x) \,
\eeq
\beq
V'(Q)_{n,n-1}  = {n\over \gamma_n} = (\beta_{n}-x) R_{n,n-1}(x)  + \gamma_{n} R_{n-1,n-1}(x) + \gamma_{n+1} R_{n+1,n-1}(x) 
\eeq
Using the symmetry, $R_{n-1,n+1}=R_{n+1,n-1}$ can be eliminated:
\beq
1= (\beta_{n}-x) (\gamma_{n+1} R_{n,n+1}(x) - \gamma_{n}  R_{n,n-1}(x) )  + \gamma^2_{n+1} R_{n+1,n+1}(x)- \gamma_{n}^2 R_{n-1,n-1}(x) 
\eeq
We have thus proven that $u_n(x) = R_{n,n-1}(x)$ and $v_n(x)=R_{n,n}(x)$ satisfy the recursion relations \eq{recuva},\eq{recuvb}.
In order to complete the proof, one must verify that the expression is correct for $n=1$. This is not difficult, and we don't write it here.

In any case, we will give an alternative proof of this result in the next section, which does require the $n=1$ case.

\section{Deformations with respect to the potential}

The deformation equations determine the variation of the orthogonal polynomials when the potential $V(x)$ is varied.
In this section, we don't assume anything about the potential (which need not be a polynomial), or about the integration contour (which can be any (non-connected) path in the complex plane; we don't have to integrate by parts).

\bigskip

For any integer $k>0$, let the parameter $u_k$ be such that:
\beq
{\d \over \d u_k} V(x) =  {1\over k} x^k  \ .
\eeq
In particular if $V(x)$ is a polynomial, we have:
\beq
V(x) = \sum_{k=1}^{\deg V} {1\over k} u_k x^k \ . 
\eeq

Expand the variation of $\psi_n(x)$ with respect to $u_k$ on the basis of $\left\{ \psi_{n}\right\}$:
\beq\label{dpsidukinfty}
{\d \over \d u_k} \psi_n(x) = \sum_{m=0}^\infty \left(U_k\right)_{nm} \psi_m(x) 
\ .
\eeq
The infinite matrix $U_k$ is antisymmetric (by differentiating \eq{ortho}):
\beq\label{Ukantisym}
(U_k)_{n,m} = - (U_k)_{m,n} \ .
\eeq
It is easy to see that (take the derivative of \eq{defpsin} w.r.t $u_k$, use $\d p_n/\d u_k = O(x^{n-1})$ and use \eq{Ukantisym}):
\beq\label{Ukdef}
U_k = -{1\over 2k} \left( \left(Q^k\right)_+ - \left(Q^k\right)_- \right)  \ .
\eeq
This implies that the infinite matrix $U_k$ is finite band: $(U_k)_{n,m} = 0$ if $|n-m|>k$, and this means that \eq{dpsidukinfty} is a finite sum.
Notice that the diagonal part yields (from \eq{defpsin}):
\beq
{1\over k} \left(Q^k\right)_{n,n} = {\d \ln{h_n} \over \d u_k} \ .
\eeq
$U_k$ can be used to form a differential system (by expressing any $\psi_{m}$ in terms of $\psi_n$ and $\psi_{n-1}$ with \eq{multx}):
\beq
{\d \over \d u_k} \pmatrix{\psi_{n-1}(x) \cr \psi_n(x)} = {\cal U}_k(x) \pmatrix{\psi_{n-1}(x) \cr \psi_n(x)} \ ,
\eeq
where ${\cal U}_k(x)$ is a $2\times 2$ matrix with polynomial coefficients of degree $\leq k$.
For instance it is easy to compute ${\cal U}_1(x)$:
\beq
{\cal U}_1(x) = -{1\over 2} \pmatrix{ \beta_{n-1}-x & 2\gamma_n \cr -2\gamma_n & x-\beta_n}  \ .
\eeq

\begin{theorem} \label{thdeform}

The expression for ${\cal U}_k(x)$ is:
\beq
{\cal U}_{k}(x) = {1\over 2k} \pmatrix{ x^k-Q^k_{n-1,n-1}  & 0 \cr 0 & Q^k_{n,n}-x^k } 
+{1\over k}\gamma_n \pmatrix{    \left({x^{k}-Q^{k}\over x-Q}\right)_{n,n-1}  &  - \left({x^{k}-Q^{k}\over x-Q}\right)_{n-1,n-1}\cr \left({x^{k}-Q^{k}\over x-Q}\right)_{n,n} &   -\left({x^{k}-Q^{k}\over x-Q}\right)_{n,n-1}  }
\eeq
Notice that:
\beq
\tr  {\cal U}_{k}(x) = {\d\over \d u_k} \ln{\gamma_n} 
\eeq
\end{theorem}

{\noindent \bf Proof:}

Notation: for any infinite matrix $A$, let $A_l$ denote the $l^{\rm th}$ diagonal above (or below, if $l<0$) the main diagonal. $A_+$ (resp. $A_-$) denotes the strictly upper (resp. strictly lower) triangular part of $A$.

\medskip

Since the matrix $Q=Q_{-1}+Q_0+Q_1$ has only three non-vanishing diagonals, we have:
\bea
\left(Q^{k+1}\right)_+ & = & (Q Q^k)_+ = (Q_1 Q^k)_+ + (Q_0 Q^k)_+ + (Q_{-1} Q^k)_+  \cr
& = & Q_1 \left(\left(Q^k\right)_+ + \left(Q^k\right)_0\right)  + Q_0 \left(Q^k\right)_+ + Q_{-1} \left(\left(Q^k\right)_+ - \left(Q^k\right)_1\right) \cr
& = & Q \left(Q^k\right)_+ + Q_1 \left(Q^k\right)_0 - Q_{-1} \left(Q^k\right)_1
 \ .
\eea
This combined with a similar calculation for $\left(Q^{k+1}\right)_-$ implies:
\beq
Q^{k+1}_+ - Q^{k+1}_- = Q (Q^k_+ -  Q^k_-) + Q_1 (Q^k_0+Q^k_{-1})  - Q_{-1} (Q^k_{1} + Q^k_0) \ ,
\eeq
which, componentwise, reads (using \eq{Ukdef}):
\bea
-2(k+1) {\d \over \d u_{k+1}} \psi_n(x) & = &  -2k {\d \over \d u_{k}} x\psi_n(x)  + \left( Q^k\right)_{n,n}( \gamma_{n+1}\psi_{n+1} -\gamma_n\psi_{n-1}) \cr
& & + \left( \gamma_n\left( Q^k\right)_{n,n-1}  - \gamma_{n+1}  \left( Q^k\right)_{n,n+1} \right) \psi_n  \ .
\eea
Using \eq{multx}:
\bea
-2(k+1) {\d \over \d u_{k+1}} \psi_n(x) & = & -2k {\d \over \d u_{k}} x\psi_n(x)  \cr
& & + \left( Q^k\right)_{n,n}( (x-\beta_n)  \psi_{n} -  2\gamma_n \psi_{n-1}) \cr
& & + \left(\gamma_n \left( Q^k\right)_{n,n-1}  - \gamma_{n+1}  \left( Q^k\right)_{n,n+1} \right) \psi_n \cr
-2(k+1) {\d \over \d u_{k+1}} \psi_{n-1}(x) & = & -2k {\d \over \d u_{k}} x\psi_{n-1}(x) \cr
& & + \left( Q^k\right)_{n-1,n-1}( 2\gamma_{n}  \psi_{n} -  (x-\beta_{n-1}) \psi_{n-1}) \cr
& & + \left(\gamma_{n-1} \left( Q^k\right)_{n-1,n-2}  - \gamma_{n}  \left( Q^k\right)_{n,n-1} \right) \psi_{n-1} \ ,
\eea
which can be further simplified using
\bea
Q^{k+1}_{n,n} & = & \gamma_{n+1} Q^k_{n,n+1} + \beta_n Q^k_{n,n} + \gamma_n Q^k_{n,n-1} \cr
Q^{k+1}_{n-1,n-1} & = & \gamma_{n-1} Q^k_{n-1,n-2} + \beta_{n-1} Q^k_{n-1,n-1} + \gamma_n Q^k_{n,n-1} \ .
\eea
We thus get a recursion formula for ${\cal U}_{k}(x)$:
$$
-2(k+1){\cal U}_{k+1}(x)  + 2k x {\cal U}_k(x)  = 
$$
\beq
\pmatrix{- \left(xQ^k-Q^{k+1}\right)_{n-1,n-1} & 0 \cr
  0 & \left(xQ^k - Q^{k+1}\right)_{n,n} }
+ 2\gamma_n
\pmatrix{ - Q^k_{n-1,n}  & Q^k_{n-1,n-1} \cr -Q^k_{n,n} & Q^k_{n,n-1} }
\eeq
from which the result follows easily.

\subsection{Alternative proof of Theorem \ref{thDeriv} }

When $V(x)$ is a polynomial, \eq{PV} implies that:
\beq
P = \sum_{k=0}^{\deg V'} u_{k+1} k U_k 
\eeq
and therefore
\beq
\DD_n(x) =  \sum_{k=1}^{\deg V'} u_{k+1} k {\cal U}_k(x) \ .
\eeq
Using theorem \ref{thdeform}, as well as \eq{eqmvt} one immediately gets the proof of theorem \ref{thDeriv}.

\section{Conclusion and generalizations}

We have found a general expression of the linear differential system $\DD_n(x)$, which is very convenient for explicit computations.

The result was derived only for a polynomial $V$, with even degree and leading coefficient positive.
It extends immediately to $V$ of arbitrary degree and leading coefficient (but now the integration path is taken in the complex plane such that the integrals converge), provided that the orthogonal polynomials exist in that case.

It could also be extended to a potential $V$ whose derivative $V'$ is a rational function. The only difference is that $V'(Q)$ is no longer a finite band matrix, and computing $\left((Q-a)^{-1}\right)_{n,n-1}$ or $\left((Q-a)^{-1}\right)_{n,n}$ is not easy (other approaches like Geronimus transformations might be more efficient \cite{Geronimus}).

Another generalization is when the integration path is not from $-\infty$ to $\infty$, but has endpoints. Again, a differential system can be found, which has poles at the endpoints. The part without poles is still given by \ref{thDeriv}, but the residues of the poles are not easily found with that method.
Other approaches (such as the dressing method of Zakharov Shabat \cite{ZakCh, Faddeev}) are more efficient.

Another generalization is for biorthogonal polynomials.
Theorems \ref{thDeriv} and \ref{thdeform} can be generalized, and the expressions will be given in a forthcoming article \cite{BEH2}.

\vfill\eject

\end{document}